\input harvmac
\def\dslash{\not{\hbox{\kern-2pt $\partial$}}}
\def\Dslash{\not{\hbox{\kern-4pt $D$}}}
\def\Oslash{\not{\hbox{\kern-4pt $O$}}}
\def\Qslash{\not{\hbox{\kern-4pt $Q$}}}
\def\pslash{\not{\hbox{\kern-2.3pt $p$}}}
\def\kslash{\not{\hbox{\kern-2.3pt $k$}}}
\def\qslash{\not{\hbox{\kern-2.3pt $q$}}}
 \newtoks\slashfraction
 \slashfraction={.13}
 \def\slash#1{\setbox0\hbox{$ #1 $}
 \setbox0\hbox to \the\slashfraction\wd0{\hss \box0}/\box0 }

 
\def\IR{\relax{\rm I\kern-.18em R}}
\def\IZ{\relax\ifmmode\hbox{Z\kern-.4em Z}\else{Z\kern-.4em Z}\fi}

\def\BR{\IR}

\def\CM{{\cal M}}

\def\bS{{\bf S}}

%

\Title{hep-th/9705057 SCIPP 97/12, RU-97-31}
{\vbox{\centerline{Comments on Higher Derivative Operators}
\centerline{in Some SUSY Field Theories}
}}
\bigskip
\centerline{Michael Dine}
\smallskip
\centerline{\it Santa Cruz Institute for Particle Physics}
\centerline{\it University of California, Santa Cruz, CA   95064}
\bigskip
\centerline{and}
\bigskip
\centerline{Nathan Seiberg}
\smallskip
\smallskip
\centerline{Department of Physics and Astronomy}
\centerline{Rutgers University}
\centerline{Piscataway, NJ 08855-0849}
\bigskip
\baselineskip 18pt
\noindent
We study the leading irrelevant operators along the flat
directions of certain supersymmetric theories.  In particular, we
focus on finite $N=2$ (including $N=4$) supersymmetric field theories in
four dimensions and show that these operators are completely determined
by the symmetries of the problem.  This shows that they
are generated only at one loop and are not renormalized beyond this
order.  An instanton computation in similar three dimensional theories
shows that these terms are renormalized.  Hence, the four dimensional
non-renormalization theorem of these terms is not valid in three
dimensions. 

\Date{5/97}

\newsec{Introduction}

Recent advances in the understanding of supersymmetric field theories
and string theory were made possible by the special properties of these
theories.  In particular, the objects which are
annihilated by some of the supersymmetries are very special and can
often be determined explicitly.  

In theories with four supercharges
(like the $N=1$ theory in four dimensions) the superpotential is
annihilated by half the supercharges and therefore it is supersymmetric
after integrating over the other half of superspace.  This property is
related to the holomorphy of the superpotential.  This fact underlies
the standard perturbative non-renormalization theorem
\ref\nonren{N. Seiberg,  ``Naturalness Versus Supersymmetric
Non-renormalization Theorems'' Phys.Lett. {\bf B318} (1993) 469,
hep-ph/9309335.}, 
and often permits an exact, non-perturbative
determination of the superpotential (see, e.g.
\ref\powerholo{``The Power of Holomorphy, Exact Results
in 4-D SUSY Field Theories,'' Talk
given at Particles, Strings, and Cosmology (PASCOS 1994), Syracuse, New
York, 19-24 May, 1994, hep-th/9408013}).
The Kahler potential in these theories is not annihilated by half the
supercharges and therefore it is in general complicated and it is not
known how to determine it exactly.

In theories with eight supercharges (like the $N=2$ theory in four
dimensions) the superpotential is trivial.  Here the Kahler potential is
determined by a holomorphic prepotential $\CF$ which is annihilated by
half the  supercharges.  The leading terms in the action are given by an
integral  over the remaining half of superspace
\ref\ntwosuperspace{B. de Wit, M.T. Grisaru and M. Rocek, Phys. Lett.
{\bf B374} (1996 297, hep-th/9601115; S.J. Gates,
M.T. Grisaru, M. Rocek and W. Siegel, {\it Superspace}
(Benjamin-Cummings 1983).} 
\eqn\ntwoaction{\int d^4 \theta \CF.}
Again, the holomorphy of $\CF$ is the key fact which makes it possible
to determine it exactly
\nref\swo{N. Seiberg and E. Witten, ``Electric-Magnetic Duality,
Monopole Condensation, and Confinement in $N=2$ Supersymmetric Yang-Mills
Theory,'' Nucl.Phys. {\bf B426} (1994) 19, hep-th/9407087.}%
\nref\swt{N. Seiberg and E. Witten,
``Monopoles, Duality and Chiral Symmetry Breaking in N=2 Supersymmetric
QCD,'' Nucl.Phys. {\bf B431} (1994) 484, hep-th/9408099.}%
\refs{\swo, \swt}.
Higher dimension operators in these theories, which are integrated over
all of superspace are not controlled by such holomorphy and are in
general hard to calculate.  For a discussion of such terms see
\nref\highdim{M. Henningson, Nucl. Phys. {\bf B458} (1996) 445,
hep-th/9507135.}%
\nref\rocek{F. Gonzalez-Rey, U. Lindstrom, M. Rocek, and R. von Unge,
``On N=2 Low Energy Effective Actions,'' Phys. Lett.
{\bf B388} (1996) 581, hep-th/9607089}%
\refs{\highdim,\rocek}.  In the next section we will show that in
the special case of the finite $N=2$ $SU(2)$ theories these terms can be
determined and they are given exactly by the one loop contribution.

In theories with sixteen supercharges (like the $N=4$ theory in four
dimensions) there is no simple superspace formalism with a finite number
of fields.  (A survey of these theories will appear in
\ref\sixteens{N. Seiberg, ``Notes on theories with 16 supercharges,''
RU-97-7, to appear.}.)
Therefore, it is not easy to find the most general
supersymmetric action.  However, a simple scaling argument (see, e.g.
\ref\gsw{M. Green, J. Schwarz and E. Witten, {\it Superstring
Theory}, Cambridge University Press, New York, 1987.})
allows us to organize them.  In any number of dimensions we can assign
weight -1 to the coordinates $X^\mu$ and hence weight 1 to the
derivatives.  Even if no superspace formalism is known, we should assign
weight $\half$ to the supercharges.  Since we plan to study functions on
the moduli space, we should assign weight zero to all the scalar moduli.
By supersymmetry, this means that we should assign weight $\half$ to all
the fermions and weight zero to all gauge fields.  If we use auxiliary
fields we assign them higher weights.  

With this assignment the superpotential in $N=1$ theories is integrated
over two $\theta$'s and leads to terms of weight one.  The Kahler
potential is integrated over four $\theta$'s and leads to terms of weight
two.  After integrating out the auxiliary fields, terms of lower weight
can be generated; e.g.\ a potential of weight zero is generated after
integrating out the auxiliary fields.

It is important to stress that this scaling is not the same as
dimensional analysis which is important in the question of
renormalizability.  For instance, the superpotential leads to terms with
two fermions and any number of scalars all of which are of weight one.
The Kahler potential leads to various terms with derivatives and if it
is not canonical, it can also lead to terms with four fermions and any
number of scalars.  All these terms are of weight two.

Using this scaling, we expect that the terms which can be controlled in
theories with sixteen supercharges are of weight four.  This would
correspond to an integral over half of superspace, if there were a
superspace formulation of the theory.  Among these are terms with four
space-time derivatives and any number of bosons.  They also include
terms with two space-time derivatives, any number of bosons and four
fermions, and terms with eight fermions and any number of bosons.  The
goal of this paper is to analyze such terms in three and four
dimensions.

In section 3 we analyze these terms in the four dimensional $N=4$
theory.  We argue that they are given exactly by the one loop
contribution.  We study their potential renormalization by instantons
and conclude that instantons cannot renormalize them.

In section 4 we study the $N=8$ theory in three dimensions where we show
that instantons do renormalize these terms.  After the completion of
this work we received two interesting papers
\nref\polpou{J. Polchinski and P. Pouliot,  ``Membrane Scattering with M
Momentum Transfer,'' NSF-ITP-97-27,hep-th/9704029.}%
\nref\dkm{N. Dorey, V.V. Khoze and M.P. Mattis, ``Multi-Instantons,
Three-Dimensional Gauge Theory, and the Gauss-Bonnet-Chern Theorem,''
hep-th/9704197.}%
\refs{\polpou,\dkm},
which also analyze these terms in greater detail than we do here, but
{}from a different perspective.

Such terms with four space-time derivatives have recently figured in the
Matrix model proposal of Banks, Fischler, Shenker and Susskind
\ref\matrixmo{T. Banks, W. Fischler, S.H. Shenker and L. Susskind,
``M Theory as a Matrix Model: A Conjecture,'' hep-th/9610043.}.  (They
are also discussed in
\ref\maldacena{J. Maldacena, RU-97-30,
to appear.}.)
For agreement with eleven dimensional supergravity it was suggested that
in the quantum mechanics problem with sixteen supercharges these terms
are not renormalized beyond one loop.  The fact that we find violation
of this non-renormalization theorem in three dimensions might suggest
that it is also violated in one dimensions (quantum mechanics).  Even if
this conclusion is correct, it is not clear what consequences this
effect has on the Lorentz invariance of the theory of 
\matrixmo. 

\newsec{Four derivative terms in $N=2$ in $d=4$}

In four dimensions, $N=2$ theories for which the one loop $\beta$
function vanishes are finite, conformally invariant theories.  Example
include gauge theories with gauge group $SU(N_c)$ and $2N_c$
hypermultiplets \swt.  For these theories, we can easily prove a
non-renormalization theorem for the $F_{\mu \nu}^4$ terms.  The proof
exploits the fact that the couplings of the vector multiplet alone are
easily written in $N=2$ superspace, as well as the scale invariance of
the theory.

Consider the case of gauge group $SU(2)$.  At low energies the theory
has one vector multiplet $\Psi$.  It can be written, in $N=2$
superspace \ntwosuperspace, as
\eqn\vectormultiplet{\Psi= \Phi+ \tilde \theta^{\alpha} W_{\alpha}
+ \tilde \theta^2\overline{D}^2 \Phi^{\dagger}}
where $\Phi$ is an $N=1$ chiral field (a function of the $N=1$
$\theta$'s), $W_{\alpha}$ is the field strength multiplet (again in
$N=1$ notation), and $\tilde \theta$ represent the extra anticommuting
coordinates required by $N=2$.  The kinetic terms arise {}from
\eqn\psiaction{\int d^2 \theta d^2 \tilde \theta \tau \Psi^2}
where $\tau = {\theta \over 4 \pi} +  {2 \pi i \over g^2}$ is the gauge
coupling.  Using the scale invariance of the theory, under which $\Psi$
has dimension one, we learn that \psiaction\ remains quadratic after
quantum 
corrections are included.  This leaves open the possibility that the
bare coefficient $\tau$ is replaced by a function of $\tau$ in the
effective theory.  The analysis of \swt\ shows that no such corrections
are present if we define $\tau$ to be compatible with duality (see also
the discussion in
\ref\fourdlanl{N. Dorey, V.V. Khoze and M.P. Mattis, ``On Mass Deformed
$N=4$ Supersymmetric Yang-Mills Theory,'' hep-th/9612231.}).

Terms of weight four arise from couplings of the type
\eqn\allsuperspace{\int d^8 \theta {\cal H}(\Psi , \Psi^{\dagger}, \tau,
\tau^\dagger).}
These must respect all the symmetries of the problem.  ${\cal H}$ must
be dimensionless, while $\Psi$ has dimension one.  ${\cal H}$ must also
respect the $U(1)_R$ symmetry.  Because the theory is scale invariant,
there can be no scale (other than $\Psi$ itself).  For fixed $\tau$
there is a unique, non-trivial form permitted by the symmetries:
\eqn\hequals{{\cal H} \sim \ln ({ \Psi \over \Lambda})\ln({\Psi^\dagger
\over \Lambda})={1\over 2}\ln^2({\Psi^{\dagger} \Psi \over \Lambda^2}) +
(f(\Psi) + c.c)} 
(the last two terms do not survive integration over superspace, keeping
in mind that $\Psi$ is chiral).  The scale $\Lambda$ here is a fake; the
terms involving $\ln\Lambda$ are multiplied by chiral or antichiral
functions, which vanish when integrated over all of superspace.
Similarly, this expression respects the $U(1)_R$ symmetry, since the
change in the Lagrangian under such a transformation is an integral over
a chiral or antichiral superfield.  Note that at general point in the
moduli space, this Lagrangian is to be interpreted by expanding the
field about that point.

To determine the $\tau$ dependence of \hequals\ we follow \nonren\ and
promote $\tau$ to a background superfield.  The coupling in \psiaction\
shows that it must be in a vector superfield.  Now, the scale invariance
and $U(1)_R$ invariance discussed in the previous paragraph are ruined
unless there is no $\tau$ dependence at all.  This statement implies
that there are neither perturbative nor non-perturbative corrections.
Thus the $F_{\mu \nu}^4$
terms arise only at one loop in the finite $N=2$ theories.

\newsec{Four derivative terms in $N=4$ in $d=4$}

The discussion of the previous section can be applied
immediately to the $N=4$ theories in $d=4$.  These theories
can be described in $N=2$ language.  Each $N=4$ multiplet consists of
a vector multiplet of $N=2$ and a hypermultiplet in the adjoint
representation.  Focusing again on $SU(2)$ gauge theory, the light
degrees of freedom on the moduli space of vacua are in a single $N=4$
multiplet.  Our choice of $N=2$ decomposition of this multiplet is such
that the scalar expectation value is in the vector of $N=2$ and the
$N=2$ hypermultiplet does not have an expectation value.  Repeating the
arguments above gives again a unique form for the terms in the effective
action involving the vector fields, $\Psi$, alone
\eqn\nequalsfourh{{\cal H} \sim \ln ({\Psi \over
\Lambda})\ln({\Psi^{\dagger } \over\Lambda}).}
Just as before, one can argue that this term is not renormalized.  

It is useful to verify that the action
indeed has this structure.  The one loop computation
is a straightforward Feynman diagram calculation.
To determine the expected form of the action, one expands
\eqn\siexpansion{\Psi = v+ \Psi^{\prime},}
where $\Psi^{\prime}$ is the fluctuating field.
Then one can develop eqn.\ \nequalsfourh\ in powers of
$\Psi^{\prime}$.  This yields (dropping terms which vanish upon
integration over all of $N=2$ superspace):
\eqn\fexpansion{{\cal H} \sim {1 \over v^2} \Psi^{\dagger} \Psi -{1
\over 
2v^3}(\Psi^{\dagger} \Psi^2 + \Psi^{\dagger 2} \Psi)
+ {1\over 3 v^4}(\Psi^{\dagger} \Psi^3 + \Psi^{\dagger 3} \Psi)
+ {1 \over 4 v^4}{\Psi^{\dagger 2} \Psi^2} + {\cal O}(\Psi^5).}

This expression can readily be rewritten in $N=1$ language, and then in
terms of component fields.  Ref.\ \rocek\  gives a
general expression for the integral of ${\cal H}$, eqn.\ \allsuperspace,
in terms of $N=1$ superfields.  In particular, there are terms
\eqn\someterms{{1 \over v^2 }\int d^4 \theta
[-2 \nabla^{\alpha \dot \alpha} \Phi^* \nabla \delta_{\alpha \dot
\alpha} \Phi + 4i \bar W^{\dot \alpha} \nabla^{\alpha}_{\dot \alpha}
W_{\alpha} + {1 \over v^2} W_{\alpha}^2 \bar W_{\dot \alpha}^2
+ \dots]}
(we are using the notation of ref.\ \rocek).  The remaining $\theta$
integrals yield a variety of component interactions.  For example,
there is a four-gaugino interaction,
\eqn\fourlambda{{4 \over v^2} \lambda \sigma^{\mu} \partial_{\mu}
\lambda^* \lambda^* \bar \sigma^{\mu} \partial_{\mu} \lambda.}
Similarly, there is a two gaugino term, with coefficient
\eqn\twogaugino{{8 \over v^2} \lambda^* \partial^2 \partial_{\mu}
\lambda.}
We have verified that these (and other) terms are generated with
the correct coefficients, by performing a calculation
in the 't Hooft-Feynman gauge.  

While this calculation is completely straightforward, it is
perhaps worth describing a few of its features.  The two
gaugino term can simply be read off of the one loop
propagator, expanding to order $p^3$.  On finds
\eqn\oneloopprop{{8 \over 3}{g^2\over 16 \pi^2 M^2}p^2 \slash p}
where $M^2$ is the vector meson mass.  For the four
fermion operators, there are many terms.  One can simplify
the computation by taking a simple choice of the external
momenta; e.g.\ one can look for a term
\eqn\fourfermi{\lambda(0) \slash p \lambda^*(p) \lambda^*(0) \slash p
\lambda(-p).} 
This term is generated by the operator $\lambda \slash \partial
\lambda^* \lambda^* \slash \partial \lambda$, which is
obtained from the $\theta$ integrations from eqn.\ \someterms.
The coefficient of this operator is $g^2{8 \over 6}{1 \over 16 \pi^2
M^4}$. The relative coefficients agree with what is expected from
eqn.\ \someterms. 

It is not difficult to generalize the term \nequalsfourh\ to include the
couplings of the hypermultiplets.  One approach is to take this
expression, written in $N=1$ language, and generalize it using the
$SU(4)$ symmetry of the theory.  In fact, it is enough to use the
$SU(3)\times U(1)_R$ subgroup which is manifest in $N=1$ superspace.  In
$N=1$ language, the theory contains a vector 
multiplet (with $R=1$) and three chiral multiplets, $\Phi_i$.  These
latter transform as a triplet of the $SU(3)$ with $R=2/3$.  The $N=2$
vector multiplet contains one of these chiral fields, e.g.\ $\Phi_3$.
Writing out eqn.\ \nequalsfourh\ in $N=1$ language, it is straightforward
to generalize it so that it respects the full symmetry.

The effective action, eqn.\ \nequalsfourh, generalized as described
above to include the couplings of the hypermultiplets, generates many
terms including certain eight fermion operators with no derivatives.  If
one examines instanton effects in the theory, it might seems that these
could generate eight fermion operators at zero momentum.  An instanton in
this theory possesses $16$ fermion zero modes, before including the
effects of the Higgs fields.  Out of the 16 supercharges 8 annihilate
the classical solution and the other 8 generate zero modes.  Similarly,
out of the 16 superconformal symmetries 8 annihilate the classical
instanton configuration and 8 generate zero modes.  If one proceeds as
in instanton calculations in $N=1$ supersymmetry
\ref\ads{I. Affleck, M. Dine and N. Seiberg, ``Dynamical Supersymmetry
Breaking in Supersymmetric QCD,''  Nucl. Phys. {\bf B241} (1984) 493.}
and in $N=2$ supersymmetry
\ref\seibergee{N. Seiberg, ``Supersymmetry and Non-Perturbative Beta
Functions,'' Phys.Lett. {\bf 206B} (1988) 75.},
one can tie eight of the 16 fermion zero modes together with background
scalars.  It is easy 
to check that these terms are non-zero.  If this were the whole
story, it would violate the
non-renormalization theorem we have just proven, so there must be some
sort of cancellation.  To see how this comes about, recall that the
background Higgs fields are of order $g$.  This is indeed what justifies
instanton calculations, such as the calculation of baryon number
violation in the standard model or the superpotentials and other
quantities in supersymmetric theories \refs{\ads, \seibergee}.  But this
means that the background 
fields are the same size as the fluctuating fields, so one must also
consider the possibility of tying together the zero modes with scalar
propagators.  In $N=1$ and $N=2$ supersymmetric QCD, this is not
possible in leading 
order.  The only relevant couplings are the Yukawa couplings of the
quarks to the gauginos (of the form $\phi^* \lambda \psi$), and the
propagator $<\phi \phi>$ vanishes to lowest order.  In the present case,
however, one has not only these Yukawa couplings, but also Yukawa
couplings arising from the superpotential, of the form $\phi \psi \psi$.
As a result, there are now numerous diagrams, with various signs.  We
have not verified the cancellation explicitly, but similar cancellations
have been seen in other contexts (e.g.\  ref.\ 
\ref\fp{D. Finnell and P. Pouliot, ``Instanton Calculations Versus Exact
Results in Four-Dimensional SUSY Gauge Theories,'' Nucl.Phys. {\bf B453}
(1995) 225, hep-th/9503115})
and we expect that they will occur here as well.

As we will discuss below, there is no possibility for an analogous
cancellation in $d=3$, so we do not expect that there is an
exact non-renormalization theorem in this case.

\newsec{Four derivative terms in $N=8$ in $d=3$}

We now turn to the study of the $N=8$ supersymmetric theory in three
dimensions.  For simplicity we consider only the gauge group $SU(2)$.
The fields in the Lagrangian are three vector multiplets each of which
consists of a vector, eight fermions and seven scalars.  The
global symmetry of the problem is $Spin(7)$.  The eight supercharges
transform as a spinor $\bf 8$ of $Spin(7)$ and hence it is an
R-symmetry.  The seven scalars are in the vector $\bf 7$ and the eight
fermions in the spinor $\bf 8$ of the symmetry group.

Along the flat directions the gauge $SU(2)$ symmetry is broken to $U(1)$
and the global $Spin(7)$ symmetry is broken to $Spin(6) \cong SU(4)$.
The light fields are in one supermultiplet.  Since the long distance
theory is free, we can dualize the vector in the multiplet to a compact
scalar 
\eqn\sigmadef{\sigma \sim \sigma + 2\pi}
and hence the moduli space is eight dimensional, parametrized by
$\sigma$ and seven scalars $\vec \phi$.  It is
\eqn\modthree{\CM= {\BR^7 \times \bS^1 \over \IZ_2}}
where we mod out by $\IZ_2$ because of the Weyl group of $SU(2)$.  The
metric on $\CM$ is the obvious flat metric and has two orbifold
singularities.  The theory at the singularity $\vec \phi=\sigma=0$ is
likely to be interacting \sixteens, while at $\vec \phi =0$,
$\sigma=\pi$, the theory is free -- it is an orbifold theory
\sixteens.

The metric on the moduli space exhibits a global $U(1)$ symmetry of
shifts along the $\bS^1$ direction.  It appears as a consequence of the
Bianchi identity of the field strength, $F$, of the ``photon'' along the
flat directions.  This symmetry is not a symmetry of the microscopic
theory.  One way to see that is to note that in terms of the fundamental
degrees of freedom $F$ is a gauge invariant composite which does not
satisfy the Bianchi identity.  Alternatively, the magnetic monopoles of
the four dimensional theory are instantons in three dimensions and they
explicitly break the global $U(1)$ symmetry.  These instantons were
first studied in the theory without supersymmetry by Polyakov
\ref\polyakov{A.M. Polyakov, Nucl.Phys.{\bf B120} (1977) 429.}.
In theories with $N=2$ supersymmetry they were discussed by Affleck,
Harvey and Witten
\ref\ahw{I. Affleck, J. Harvey and E. Witten, Nucl. Phys.
{\bf B206} (1982) 413.}
and in $N=4$ in
\nref\swthree{N. Seiberg and E. Witten,``Gauge Dynamics
and Compactification to Three Dimensions,
`` IASSNS-HEP-96-78, hep-th/9607163.}
\nref\losalamosf{N. Dorey, V.V. Khoze, M.P. Mattis,
D. Tong and S. Vandoren,``Instantons,
Three-Dimensional Gauge Theory, and the Atiyah-Hitchin Manifold,''
hep-th/9703228.}%
\refs{\swthree,\losalamosf}.  More recently they were also discussed in
theories with $N=8$ supersymmetry \refs{\polpou,\dkm}. 
Here we mention only the main properties of the instanton computation.

The expectation values of the scalar fields break the gauge symmetry to
$U(1)$ and the global symmetry to $SU(4)$.  In this set up there are BPS
saturated field configurations which are monopoles in four dimensions
and instantons in three.  The instanton configuration breaks some of the
remaining unbroken symmetries.  For example, translation invariance in
the three space-time dimensions is broken.  For every such broken
generator there is a collective coordinate which should be integrated
over.  Of particular interest are the fermion zero modes.  They are also
associated with symmetries of the problem.  The sixteen supercharges can
act on the classical configuration.  Eight of them annihilate it and
therefore the instanton is BPS saturated.  The other eight supercharges
lead to fermion zero modes.  Since they are associated with symmetry
generators, one can introduce collective coordinates for them.  It is
easy to see that there are no other fermion zero modes.

The eight fermion zero modes transform as two $\bf 4$'s of the unbroken
global $SU(4)$.  They lead to an effective interaction which is a
product of eight fermions
\eqn\eightferm{ \prod_{I, A} \lambda^I_A}
where $I=1,...,4$ and $A=1,2$.  
Each fermion in this interaction transforms
as $({\bf 4, 2 })$ of $SU(4) \times SU(2)$ (the $SU(2)$ factor is the
Euclidean space Lorentz group in three dimensions).   The interaction
\eightferm\ is $SU(4) \times SU(2)$ 
invariant.  The full interaction term depends on the expectation values
of the scalar fields.  It includes a factor of 
\eqn\instfac{\exp{\left ( -{<|\vec \phi|> \over g^2 } + i \sigma
\right)} .} 
The first term in the exponent $<|\vec \phi|> \over g^2 $ is the
instanton action which depends on the scalar expectation value and the
gauge coupling $g$.  The second term $i \sigma$ shows that the instanton
breaks the global $U(1)$ of shifts of $\sigma$ by a constant.  More
factors of $<\vec \phi>$ are needed to make \eightferm\ not only $SU(4)$
invariant but $Spin(7)$ invariant.

Obviously, the full term we have just described is complex.  In the
action we have to add to it its hermitian conjugate.  It is generated by
an anti-instanton and involves the product of eight fermions
transforming as $({\bf \overline 4, 2 })$ of $SU(4) \times SU(2)$.

It is useful to compare this instanton computation to the analogous
computation in four dimensions (section 3).  In four dimensions the
theory is conformally invariant and therefore there are sixteen fermion
zero modes (eight from supersymmetry and eight from the superconformal
symmetry).   In three dimensions the theory is not conformally invariant
and hence there are only eight zero modes.  Furthermore, in three
dimensions there is an exact instanton configuration which is BPS
saturated (annihilated by half the supercharges).  This is not true in
four dimensions, where constrained instantons
\ref\const{I. Affleck, Nucl.Phys. {\bf B191} (1981) 429.}
have to be used.  This makes the computation more delicate in four
dimensions and leaves the possibility of the cancellation discussed in
section 3.

We conclude that instantons generate terms with eight fermions.
Therefore, by the discussion in the introduction they renormalize the
terms in the effective action with eight fermions and also the terms
with four derivatives.  It is possible that these terms are also
corrected by perturbative loop effects.  Whether or not this is so, the
non-renormalization theorem for these terms is not true in three
dimensions. 

\bigskip
\centerline{{\bf Acknowledgements}}
\noindent
We thank T. Banks, W. Fischler, J. Polchinski, S. Shenker, I. Singer, L.
Susskind and E. Witten for helpful discussions.  This work was supported
in part by DOE grants DE-FG02-96ER40559 
and DE-FG03-92ER40689.

\listrefs

\end